\documentclass[useAMS,usenatbib]{mn2e}
\usepackage{graphicx}

\title[Detection of a radio bridge in Abell 3667]
      {Detection of a radio bridge in Abell 3667}
\author[E. Carretti, et al.]
{E.~Carretti$^{1}$\thanks{E-mail: Ettore.Carretti@csiro.au (EC); 
                                                             brown@astro.umn.edu (SB); 
                                                             Lister.Staveley-Smith@icrar.org (LSS);
                                                             Jurek.Malarecki@icrar.org (JMM); 
                                                             gbernardi@cfa.harvard.edu (GB);
                                                              bryan.gaensler@sydney.edu.au (BMG); 
                                                             m.haverkorn@astro.ru.nl (MH); 
                                                             Michael.Kesteven@csiro.au (MJK); 
                                                             spoppi@oa-cagliari.inaf.it (SP)
                                                             },
S.~Brown$^{2,3}$,
L.~Staveley-Smith$^{4,5}$,
J.M.~Malarecki$^{4}$,
G.~Bernardi$^{6}$,
 \newauthor
B.M.~Gaensler$^{7}$,
M.~Haverkorn$^{8,9}$,
M.J.~Kesteven$^{2}$, and
S.~Poppi$^{10}$\\
$^{1}$CSIRO Astronomy and Space Science, PO Box 276, Parkes, NSW 2870, Australia\\
$^{2}$CSIRO Astronomy and Space Science, PO Box 76, Epping, NSW 1710, Australia\\
$^{3}$Department of Physics and Astronomy, University of Iowa, Iowa City, Iowa 52242\\
$^{4}$International Centre for Radio Astronomy Research, M468, University of Western Australia, Crawley, WA 6009, Australia\\
$^{5}$ARC Centre of Excellence for All-Sky Astrophysics\\
$^{6}$Harvard--Smithsonian Center for Astrophysics, 60 Garden Street, Cambridge, MA, 02138, USA\\
$^{7}$Sydney Institute for Astronomy, School of Physics A29, The University of Sydney, NSW 2006, Australia\\
$^{8}$Department of Astrophysics/IMAPP, Radboud University Nijmegen, P.O. Box 9010, 6500 GL Nijmegen, The Netherlands\\
$^{9}$Leiden Observatory, Leiden University, P.O. Box 9513, 2300 RA Leiden, The Netherlands\\
$^{10}$INAF Ð Osservatorio Astronomico di Cagliari, St. 54 Loc. Poggio dei Pini, I-09012 Capoterra (CA), Italy}

\begin{document}

\date{Accepted xx xx xx. Received yy yy yy; in original form zz zz zz}

\pagerange{\pageref{firstpage}--\pageref{lastpage}} \pubyear{2013}

\maketitle

\label{firstpage}

\begin{abstract}
We have detected a radio bridge of unpolarized synchrotron emission 
connecting the NW relic
of the galaxy cluster Abell 3667 to its central regions. We used data at 2.3~GHz from the S-band Polarization All Sky Survey (S-PASS) and at 3.3~GHz from a follow up observation, both conducted with the Parkes Radio Telescope. This emission is further aligned with a diffuse X-ray tail, and represents the most compelling evidence for an association between intracluster medium turbulence and diffuse synchrotron emission. This is the first clear detection of a bridge associated both with an outlying cluster relic and X-ray diffuse emission. 
All the indicators point toward the synchrotron bridge being related to the post--shock  turbulent wake trailing the shock front generated by a major merger in a massive cluster. 
Although predicted by simulations, this is the first time such emission is detected with high significance and clearly associated with the path of a confirmed shock.
Although the origin of the relativistic electrons is still unknown, the turbulent re-acceleration model provides a natural explanation for the large-scale emission. 
The equipartition magnetic field intensity of the bridge is $B_{\rm eq} = 2.2\pm0.3$~$\mu$G. 
We further detect diffuse emission coincident with the central regions of the cluster for the first time. 
\end{abstract}


\section{Introduction}\label{introSec}

A number of galaxy clusters exhibit large scale diffuse emission up to a few Mpc scale either from the central region (radio halos) or in the periphery (radio relics). Not directly associated with the activity of individual galaxies, these sources are characterized by low-surface brightness ($\sim 1 \mu$Jy/arcsec$^2$ at 1.4 GHz) and steep-spectrum\footnote{$S(\nu)\propto \nu^{\alpha}$, with $\alpha$=spectral index} ($\alpha<-1$) and are proof of the existence of relativistic electrons and magnetic fields spread on large scales in the  intracluster medium (ICM) (e.g., \citealt{feretti05,ferrari08,feretti12}). These large scale structures are related to other cluster properties in the optical and X-ray domain, and are directly connected to the cluster history and evolution \citep{cassano10}.
Observations of the synchrotron emission shows that the radio halos are mostly unpolarized, while the peripheral relics are highly polarized (e.g., \citealt{feretti05}).

The mechanism that (re-)accelerates the relativistic electrons is still debated, but it must occur in situ because of their short radiative lifetimes. Proposed origins of radio halos include {\it primary} acceleration of electrons due to turbulence (e.g. \citealt{brunetti01,petrosian01}) or {\it secondary} production of electron/positron pairs (CRe$^{\pm}$) due to hadronic collisions between cosmic-ray protons (CRp) and thermal protons (e.g. \citealt{blasi99,pfrommer08}). The former requires some dynamical activity like cluster mergers to generate turbulent electron re--acceleration, and is expected to turn off quickly as electrons lose energy. The latter is expected to be ubiquitous in the cluster and continuously produced lasting a cluster lifetime, as cosmic ray protons live longer than a Hubble time.

The radio relics are instead believed to be generated by shocks propagating to the cluster outskirts after a major merger \citep{ensslin01}, though claims have been made that some relics are caused by matter infalling from Cosmic Web filaments (e.g., \citealt{ensslin98,brown11}).

In several clusters, there have also been detections of a ``bridge" connecting a peripheral relic to the central halo. 
The Coma cluster, Abell clusters 512, 1300, 2255, 2744, and the cluster 1RXS~J0603.3+4214 all show such bridges
\citep{kim89,dallacasa09,venturi12,pizzo09,orru07,vanweeren12}. The origin of relic-halo bridges is unknown, though they are clearly related to the mechanism creating the relics. Due to their large extent, however, in situ acceleration must still be responsible for the emission. 
The case of 1RXS~J0603.3+4214 has been suggested to be related to the turbulence behind the relic \citep{vanweeren12}.

The galaxy cluster Abell 3667 (A3667) is an ideal laboratory to study the physics of the ICM. It is a double cluster undergoing a merger where the secondary cluster is moving at high speed mostly in the plane of the sky \citep{owers09}. Its redshift of 0.0556 locates it a distance of 224~Mpc. Interferometric radio observations show double relics in the NW and SE outlaying regions, though no radio-halo was previously detected coincident with the diffuse X-rays of the larger sub-cluster  \citep{rottgering97, johnstonhollitt03}. The observational suite of radio data is still incomplete, however. Single-dish observations, essential to show the whole extent of the diffuse emission, have not been available. 

Using X-ray, optical, and radio interferometric data, \citet{finoguenov10} interpret this system as two post merger clusters that have generated the two relics as outgoing front shocks with the secondary member heading away from the centre and following the NW shock. If this interpretation is correct, the post--shock region behind the relics are expected to generate turbulence \citep{kang07,paul11,vazza11}, though there is currently no direct evidence for this phenomenon.

In this paper we report the detection of new large--scale synchrotron emission in the galaxy cluster A3667 previously missed by interferometric observations (e.g. \citealt{rottgering97}). Using 2.3~GHz data of the S-band Polarization All Sky Survey (S-PASS) taken with the Parkes radio telescope and a follow-up at 3.3~GHz with the same telescope, we detect a radio bridge connecting the A3667 NW relic to the central regions of the cluster. Diffuse emission from the
cluster centre is also detected.
The observations are described in Section~\ref{obs:Sec}, results and maps in Section~\ref{maps:Sec}. Comparison of our maps with radio interferometric, optical, and X-ray data 
are discussed in Section~\ref{disc:Sec} along with our interpretation of the new overall picture.

We assume  
$H_0 = 73$~km/s/Mpc, $\Omega_m = 0.27$, and $\Omega_\Lambda = 0.73$
as cosmological parameters throughout the paper.

\section{Observations of A3667}\label{obs:Sec}
\subsection{Parkes 2.3 GHz}
The S-band Polarization All Sky Survey (S-PASS) is a single-dish 
survey of the total intensity and polarized continuum emission of the entire southern 
sky at 2.3~GHz. The observations have been conducted with the Parkes Radio Telescope, NSW Australia,
a 64-m telescope operated as National Facility by ATNF-CASS a division of CSIRO.
A description of S-PASS observations and analysis 
is given in \citet{carretti11} and Carretti et al. (2012, in preparation). 
Here we report a summary of the main details. 

S-PASS observations are centred at the frequency of 2300~MHz, 
with 256~MHz bandwidth and 512 frequency channels 0.5~MHz each. 
The standard Parkes S-band receiver ({\it Galileo}) was used with 
a system temperature of $T_{\rm sys} = 20$~K and beam width
FHWM=8.9' at 2300 MHz. This is a circular polarization system, 
ideal for Stokes Q and U measurements. 
The Digital Filter Bank Mark 3 (DFB3) was used, 
recording full Stokes products.  
Flux calibration was done with PKS~B1934-638; secondary calibration with PKS~B0407-658; and  polarization calibration with PKS~B0043-424.

Data were binned in 8~MHz channels and, after RFI flagging, 
23~subbands were used covering the ranges 
2176-2216 and 2256-2400~MHz, for an effective central frequency of
2307~MHz and bandwidth of 184~MHz. 

The scanning strategy is based on long azimuth scans taken in the east and the west 
to realise absolute polarization calibration of the data.
Final maps are convolved to a beam of FWHM=10.75'.
Stokes I, Q, and U sensitivity 
is better than 1.0~mJy/beam 
everywhere in the covered area.
Details of scanning strategy, map-making, and final maps obtained by binning 
all frequency channels are presented in Carretti et al. (2012, in preparation; 
see also \citealt{carretti11}). 
Multi--frequency maps and analysis will be presented in later papers.
The confusion limit at this frequency and resolution is 6~mJy \citep{uyaniker98} in Stokes I, and much lower in polarization 
(average polarization fraction of compact sources is lower than 2\%, \citealt{tucci04}).
 From the Galactic synchrotron emission analysis at 1.4~GHz by~\citet{laporta08} we estimate a Galactic rms emission at the Galactic latitude of A3667 ($b\sim -30^\circ$) of 8~mJy at 2.3 GHz at the beam scale.

\begin{figure*}
\centering
  \includegraphics[angle=0, width=0.49\hsize]{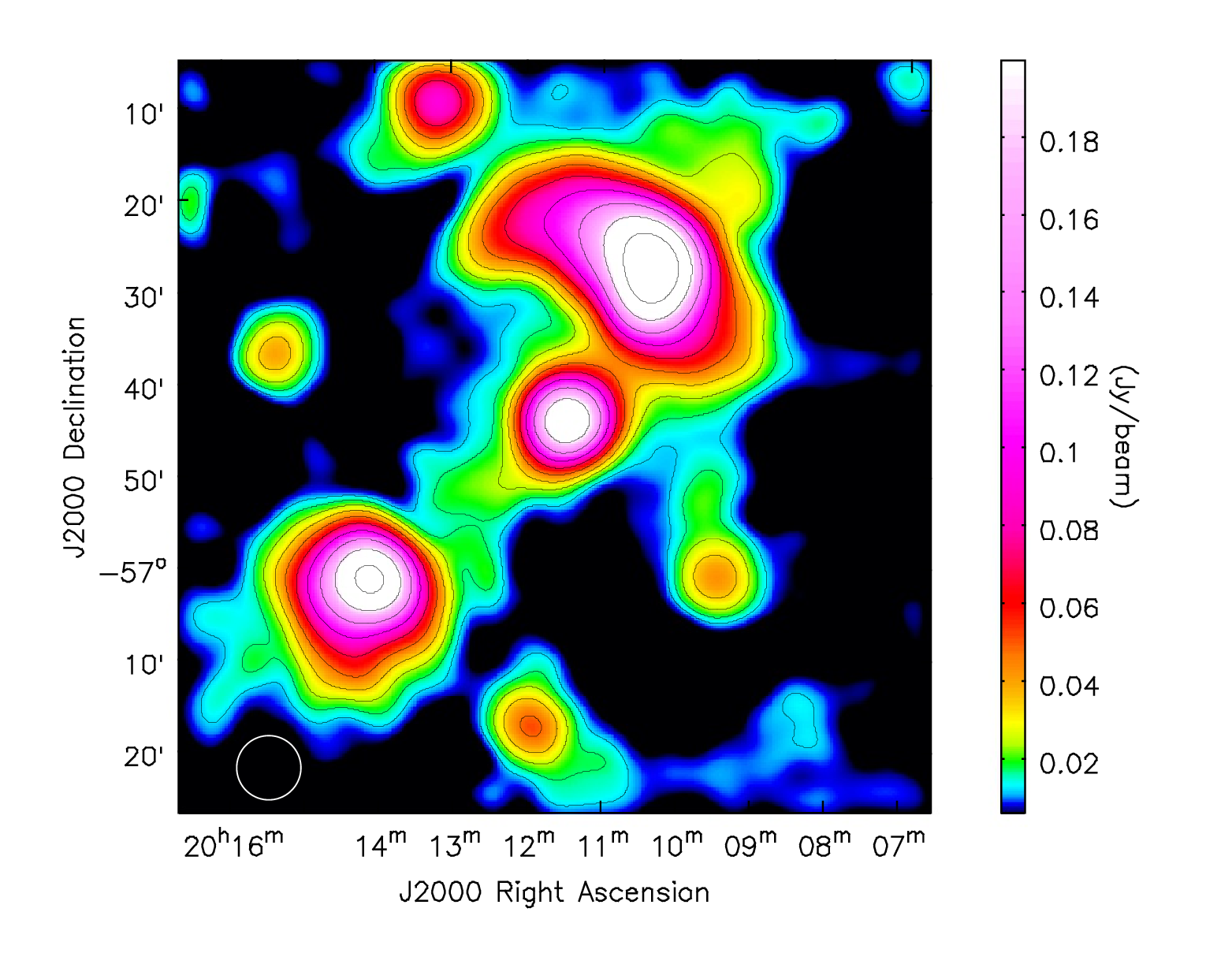}
  \includegraphics[angle=0, width=0.49\hsize]{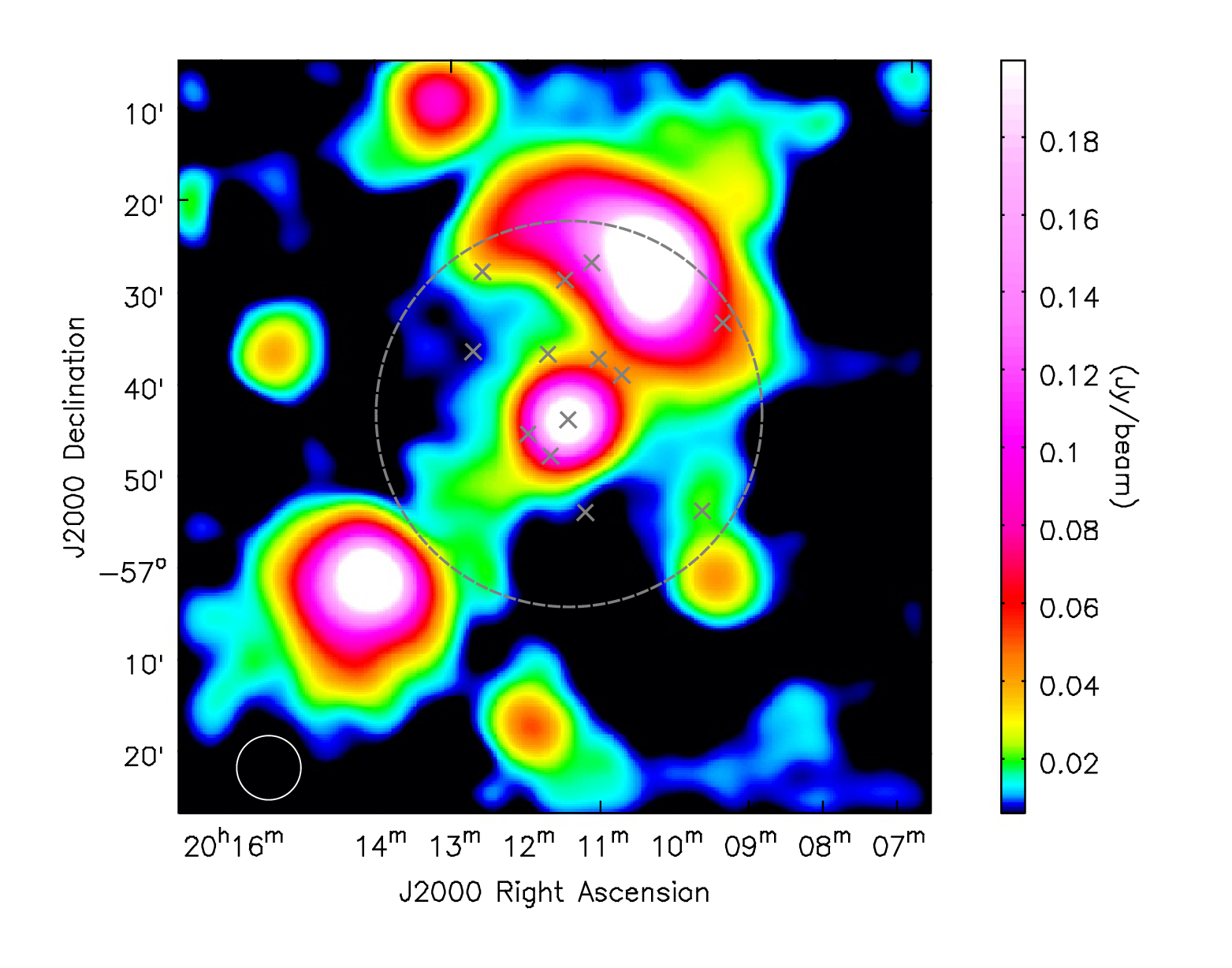}
  \includegraphics[angle=0, width=0.49\hsize]{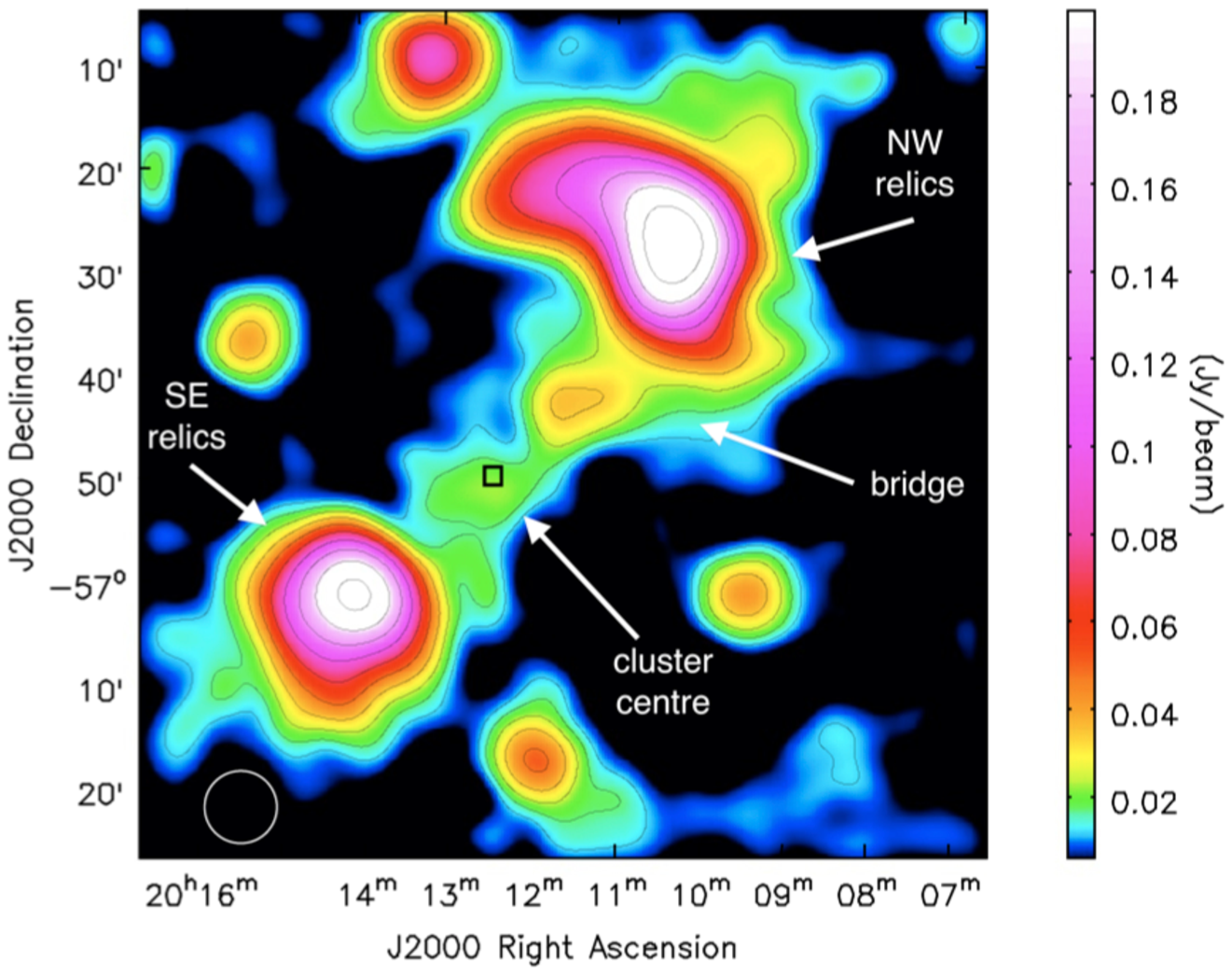}
  \includegraphics[angle=0, width=0.49\hsize]{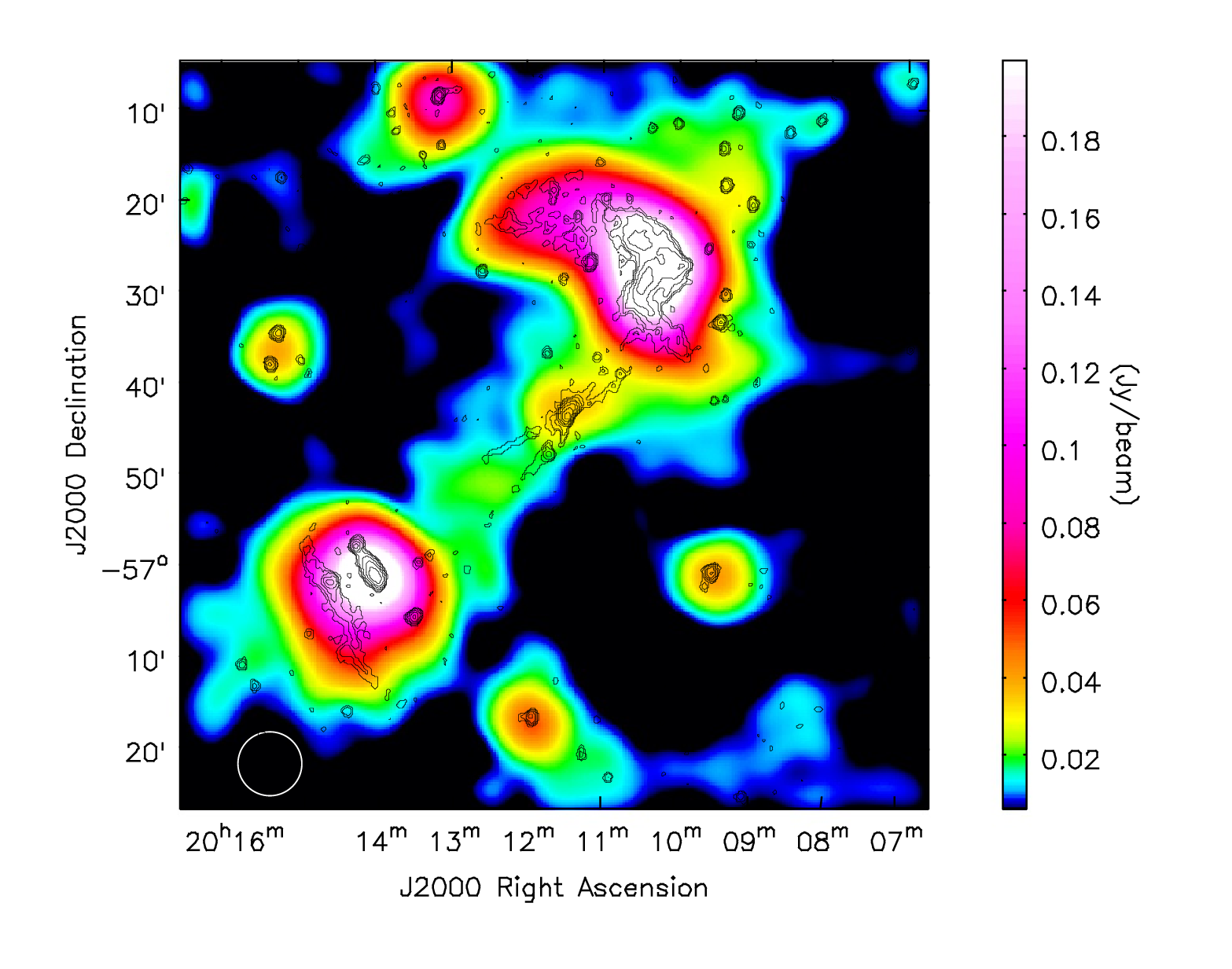}
\caption{{\bf Top--left:} Stokes I image of A3667 at 3.3 GHz taken with the Parkes radio telescope. 
                Contour levels start from 12 mJy/beam and scale by a factor $\sqrt{2}$ thereafter.
               The noise budget is dominated by the confusion limit of 3~mJy. 
               Beam size is 7.0~arcmin.
              {\bf Top--right:} Same image, but with the positions of the sources 
                detected with the ATCA observations marked (crosses). The area covered 
                by the ATCA  is also shown (gray circle).
               {\bf Bottom--left:} Same figure but with B2007-569 and the other 
                compact sources in the ATCA field subtracted. The entire area between 
                the two relics is cleaned from compact sources.
               The centre of the cluster is marked with a black square and major features
               described in the text are labelled.              
               {\bf Bottom--right:} Same source--subtracted map as bottom--left panel,
              but with the contour levels of 843 MHz
              SUMSS emission overlaid.  Levels start at 3~mJy/beam and scale 
               by a factor of 2 therafter. The beam is 43"~$\times$~43"~cosec(Dec).                 
                \label{a3667_10cm:Fig}
}
\end{figure*}

\subsection{Australia Telescope Compact Array}
To aid subtraction of point sources, particularly the head--tail galaxy B2007-569, 
interferometric observations were taken on 2011 November 26 with the Australia 
Telescope Compact Array at Narrabri. The field was observed for 9~hr with the 
telescope in its 1.5D configuration, resulting in a nominal resolution of 18 arcsec 
at 2.3 GHz. 
Data was gathered in the frequency range 1.2--3.0~GHz, but
images were made only in the S-PASS band at 2.3~GHz. 14 bands 128-MHz 
were used to analyse the frequency behaviour.
As with the S-PASS observations, 
the flux density scale was calibrated using PKS B1934-638, 
to ensure the same flux scale reference for both single-dish and interferometric observations.
Gaussian fits were made to the compact sources, and subtracted from the Parkes data as explained in Section~\ref{maps:Sec}.

\subsection{Parkes 3.3 GHz}

To get finer single--dish resolution and an independent confirmation 
of the bridge, we conducted follow up single--dish observations with the 10~cm receiver 
of the Parkes telescope on 2012 January 12 and 14 
for a total of 6~hr. The receiver is a linear polarization system, so that only Stokes I 
measurements of this data set were used.
Although the entire 2.6-3.6 GHz bandwidth
was observed, only the top 400 MHz range centred at 3.35~GHz 
was used to optimise for resolution (FWHM = 5.9'). 
The flux density scale was 
again
calibrated using PKS B1934-638.
The DFB3 backend was used with a bandwidth of 1024 MHz and 512 frequency channels 2~MHz each. All channels of the 400~MHz sub-band were binned together.  A standard basket weaving technique with orthogonal scan sets along R.A. and Dec. 
spaced by 2' was used to observe an area of $3^\circ\times3^\circ$ centred at the cluster. 
Fourier based software was applied to
make the map  \citep{carretti10}.  
This recovers the emission up to the scale of the area mapped, sufficient
for the scales covered by the cluster (up to approximately 1$^\circ$).
 Final maps are convolved to a beam of FWHM=7.0'.
The sensitivity is limited by the confusion limit (3~mJy).
 From ~\citet{laporta08} we estimate a Galactic rms emission at the A3667 latitude of $\sim2$~mJy at 3.35~GHz at the beam scale.

\section{A3667 images}\label{maps:Sec}

Figure~\ref{a3667_10cm:Fig} shows the 3.3~GHz Total Intensity 
map centred at A3667 before compact source subtraction.
The area covered with the ATCA observations and the positions of the sources found 
with them are also shown. 

\begin{figure}
\centering
  \includegraphics[angle=0, width=1.0\hsize]{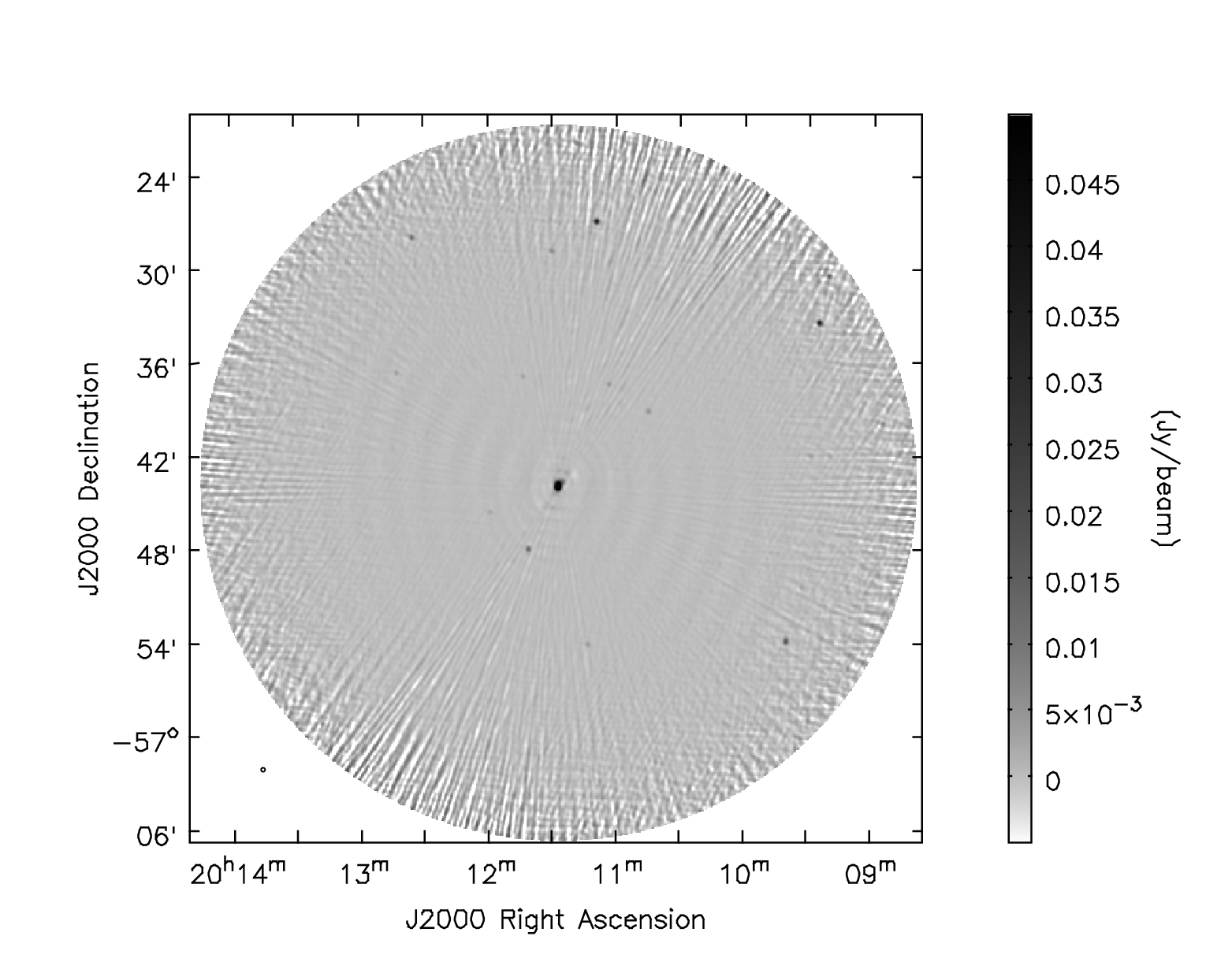}
  \includegraphics[angle=0, width=1.0\hsize]{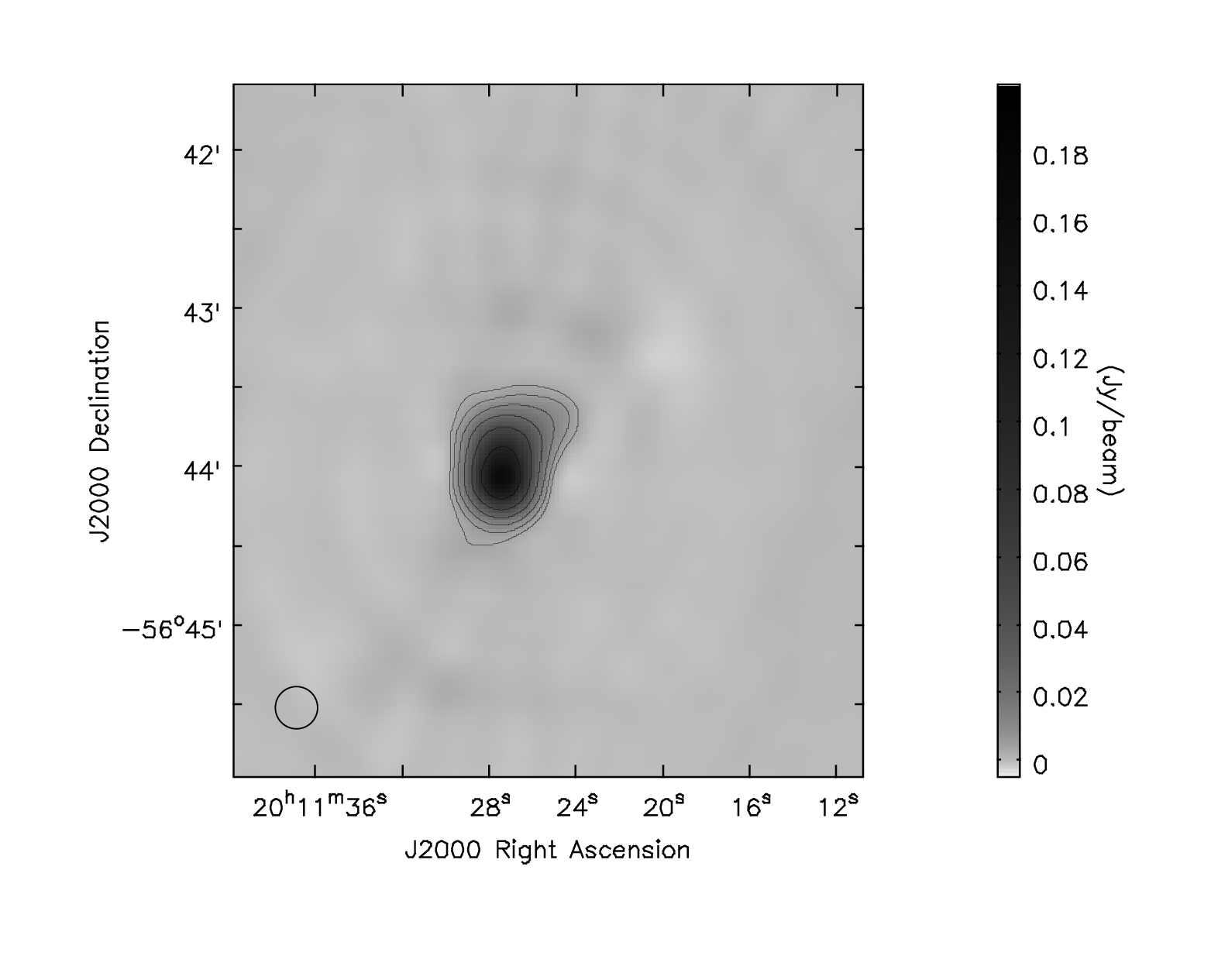}
\caption{{\bf Top:} Stokes I image of area observed with the ATCA and centred 
                                 at the head--tail radio galaxy B2007-569. This image 
                                 has been made using the same frequency
                                 coverage of the S-PASS observations.
                                 The rms noise is 0.3 mJy/beam. 
                                Beam size is 16~arcsec.
               {\bf Bottom:} Closeup of the image focused at the head--tail radio galaxy. 
                                     Contour levels start at 3~mJy/beam and scale by a factor 2.
                \label{b2007:Fig}
}
\end{figure}

\begin{figure}
\centering
  \includegraphics[angle=0, width=1.0\hsize]{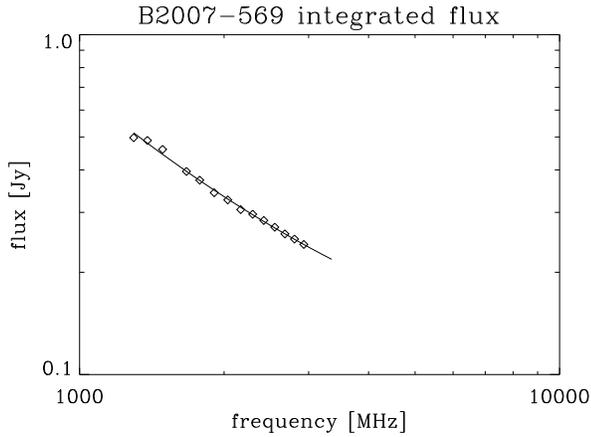}
  \caption{Integrated flux of B2007-569 (diamonds) measured in 128~MHz subbands in the 1.2--3.0 GHz range of the ATCA data. The best fitting model of the form $\log S  = A + B \log \nu + C (\log \nu)^2$ is also shown (solid line).
                \label{b2007_fit:Fig}
}
\end{figure}

An image of the ATCA data at 2.3 GHz is shown in Figure~\ref{b2007:Fig}. We find 13 sources with a 2.3~GHz flux density higher than 3~mJy. The central extended source is the head--tail radio galaxy B2007-569 (see close up in Figure~~\ref{b2007:Fig}), which is the bright spot between the two relics in the single-dish image. Its spectral behaviour in the ATCA band 1.2--3.0 GHz is shown in Figure~\ref{b2007_fit:Fig} along with the quadratic best fitting in logarithmic space 
$\log S  = A + B \log \nu + C \log^2 \nu$, where $S$ is the flux density and $\nu$ the frequency. The other 12 sources have been modelled with a simple power law $S  \propto \nu^\alpha$. 
The flux density at 2.307~GHz (measured) and 3.350~GHz (best--fitting estimate) and spectral index are reported in Table~\ref{b2007_src:Tab}. The running spectral index of B2007-569 $\alpha = d \log S / d \log \nu$  ranges from -1.0 at 1.2~GHz to -0.8 at 3.0~GHz.

Figure~\ref{a3667_10cm:Fig} also shows the Parkes 3.3~GHz total intensity 
map after source subtraction. In particular, 
a flux density of $219\pm5$ mJy\footnote{best fitting rms error.}
has been assumed for B2007-569.
Figure~\ref{a3667_10cm:Fig} also shows the same image overlaid 
with the contour levels of the emission measured
at 843~MHz by the Sydney University Molonglo Sky Survey (SUMSS, \citealt*{bock99}).

\begin{table*}
 \centering
 \begin{minipage}{200mm}
  \caption{Compact sources above 3~mJy detected in the ATCA observations.}
  \begin{tabular}{@{}lccccccl@{}}
  \hline
   ID     &  RA & Dec &   \multicolumn{2}{c}{Flux  density [mJy]}      & $\alpha$\footnote{spectral index. For B2007-569 it is a running spectral index function of the frequency $\nu$ defined as $\alpha = d \log S / d \log \nu$}   &  
   $\sigma_\alpha$\footnote{rms error on $\alpha$} & Notes \\
            &  (J2000) &  (J2000)    &  2.3~GHz \footnote{Measured} & 3.3~GHz \footnote{Best fitting estimate} &   &    & \\
 \hline
  1  &  20$^{\rm h}$11$^{\rm m}$27.5$^{\rm s}$  &  $-56^\circ$44'04''  &    280  &    219 &    $-0.90 + 0.84 \log (\frac{\nu}{2.1 \rm{GHz}})$  &   $0.021  \sqrt{ 1 + 304 \log^2(\frac{\nu}{2.1 \rm{GHz}})}$ & B2007-569 \\
  2  &  20$^{\rm h}$12$^{\rm m}$35.5$^{\rm s}$  &  $-56^\circ$27'58''  &    12.9  &    11.4 &   -0.60  &     0.15  &     \\
  3  &  20$^{\rm h}$11$^{\rm m}$ 9.4$^{\rm s}$  &  $-56^\circ$26'58''  &    36.4  &    27.8 &   -0.75  &     0.06  &     \\
  4  &  20$^{\rm h}$11$^{\rm m}$30.4$^{\rm s}$  &  $-56^\circ$28'52''  &     6.8  &     4.8 &   -0.70  &     0.22  &     \\
  5  &  20$^{\rm h}$09$^{\rm m}$25.3$^{\rm s}$  &  $-56^\circ$33'27''  &    38.1  &    35.5 &   -0.45  &     0.10  &     \\
  6  &  20$^{\rm h}$12$^{\rm m}$43.1$^{\rm s}$  &  $-56^\circ$36'41''  &     3.9  &     3.2 &   -0.76  &     0.14  &     \\
  7  &  20$^{\rm h}$11$^{\rm m}$43.8$^{\rm s}$  &  $-56^\circ$36'58''  &     3.0  &     1.7 &   -1.27  &     0.12  &     \\
  8  &  20$^{\rm h}$11$^{\rm m}$ 3.7$^{\rm s}$  &  $-56^\circ$37'28''  &     5.1  &     3.3 &   -0.62  &     0.21  &     \\
  9  &  20$^{\rm h}$10$^{\rm m}$45.1$^{\rm s}$  &  $-56^\circ$39'11''  &     5.4  &     3.6 &   -0.78  &     0.15  &     \\
 10  &  20$^{\rm h}$11$^{\rm m}$59.2$^{\rm s}$  &  $-56^\circ$45'38''  &     3.1  &     2.3 &   -0.48  &     0.19  &     \\
 11  &  20$^{\rm h}$11$^{\rm m}$41.3$^{\rm s}$  &  $-56^\circ$48'02''  &    14.4  &     9.5 &   -0.73  &     0.05  &     \\
 12  &  20$^{\rm h}$11$^{\rm m}$13.4$^{\rm s}$  &  $-56^\circ$54'11''  &     4.5  &     3.8 &   -0.69  &     0.13  &     \\
 13  &  20$^{\rm h}$09$^{\rm m}$40.2$^{\rm s}$  &  $-56^\circ$53'56''  &    20.0  &    13.0 &   -0.84  &     0.12  &     \\
\hline
\label{b2007_src:Tab}
\end{tabular}
\end{minipage}
\end{table*}
\begin{figure}
\centering
  \includegraphics[angle=0, width=1.0\hsize]{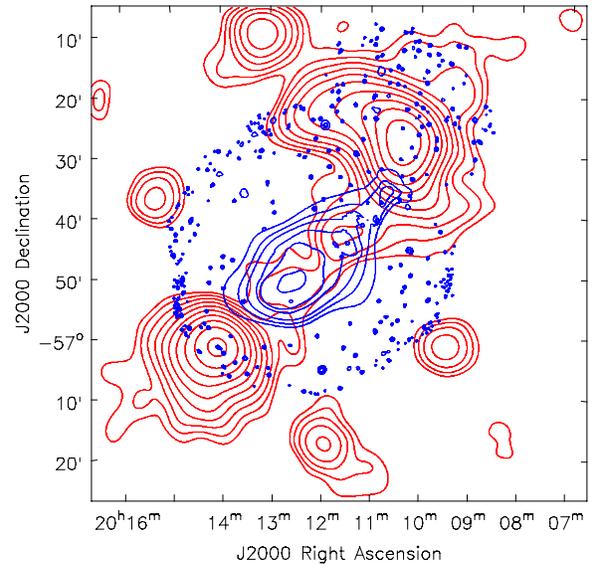}
\caption{Contours levels of the Stokes I image of A3667 at 3.3~GHz (red) over the XMM X-ray image (blue contour levels), showing the southern extension coincident with the cluster centre.  
X-ray levels are  1.6, 6.4, 12.5, 25, 50, 200x10$^{-6}$~cts/s/arcmin$^2$.
                \label{a3667_10cm_xray:Fig}
}
\end{figure}
\begin{figure*}
\centering
  \includegraphics[angle=0, width=0.49\hsize]{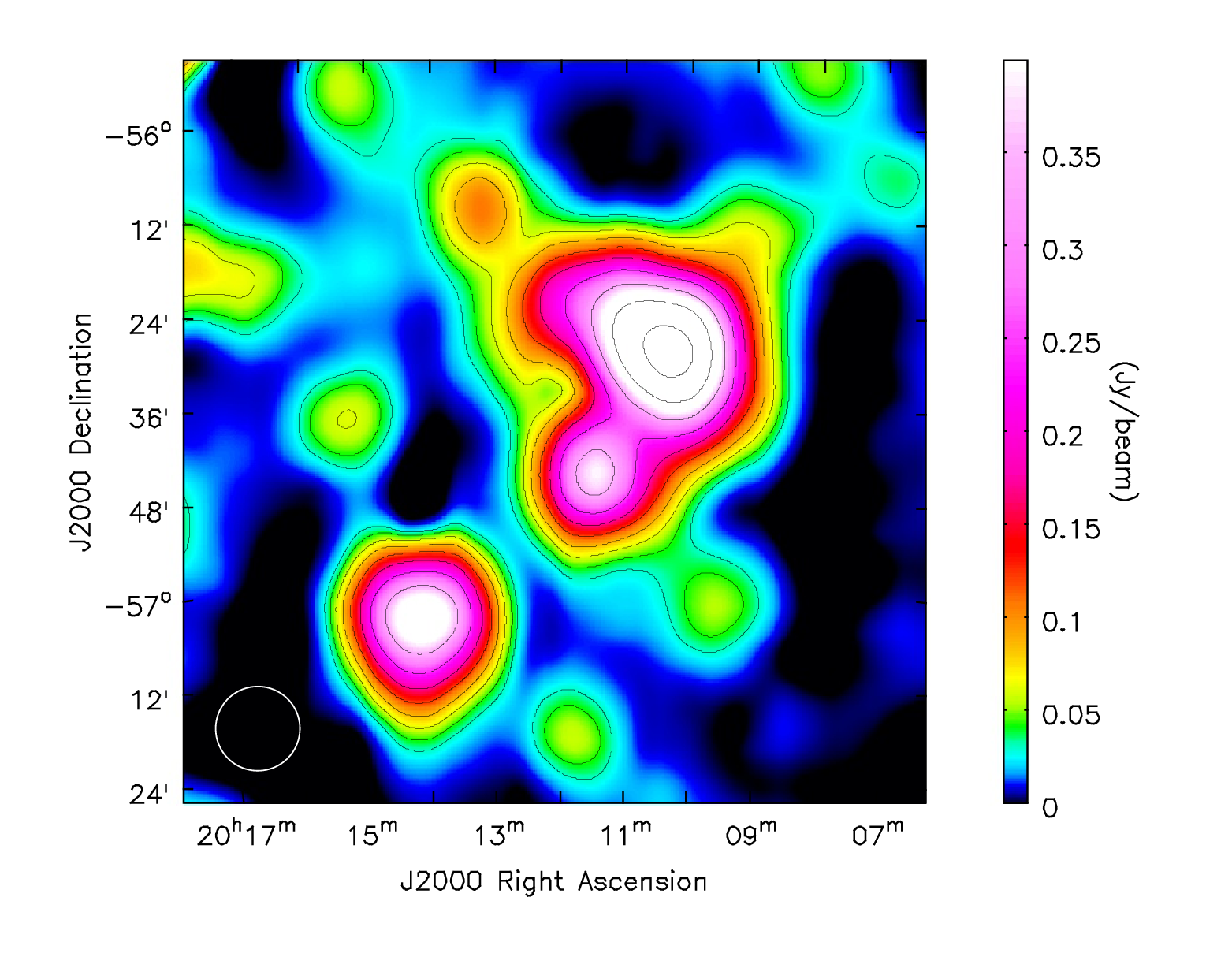}
  \includegraphics[angle=0, width=0.49\hsize]{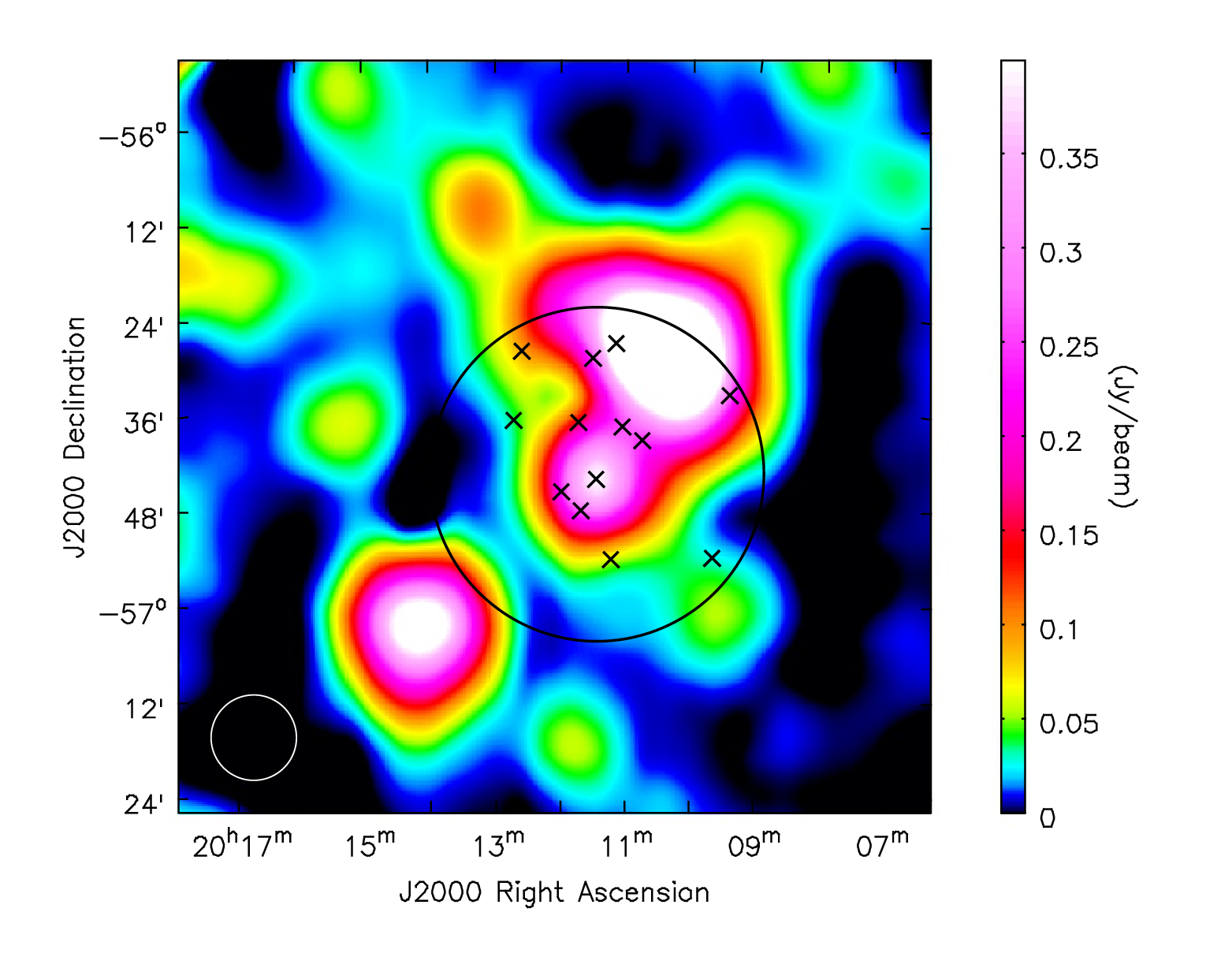}
  \includegraphics[angle=0, width=0.49\hsize]{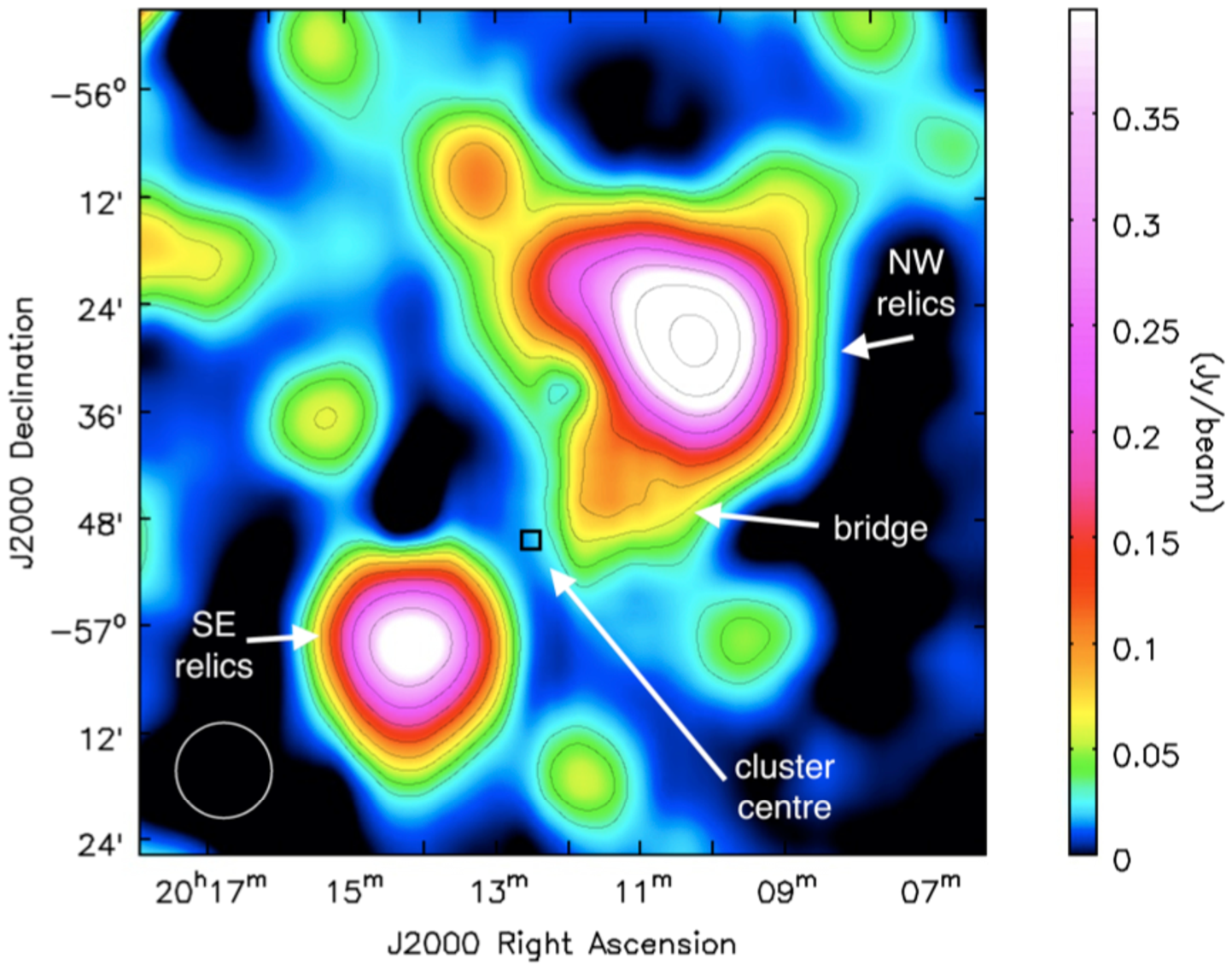}
  \includegraphics[angle=0, width=0.49\hsize]{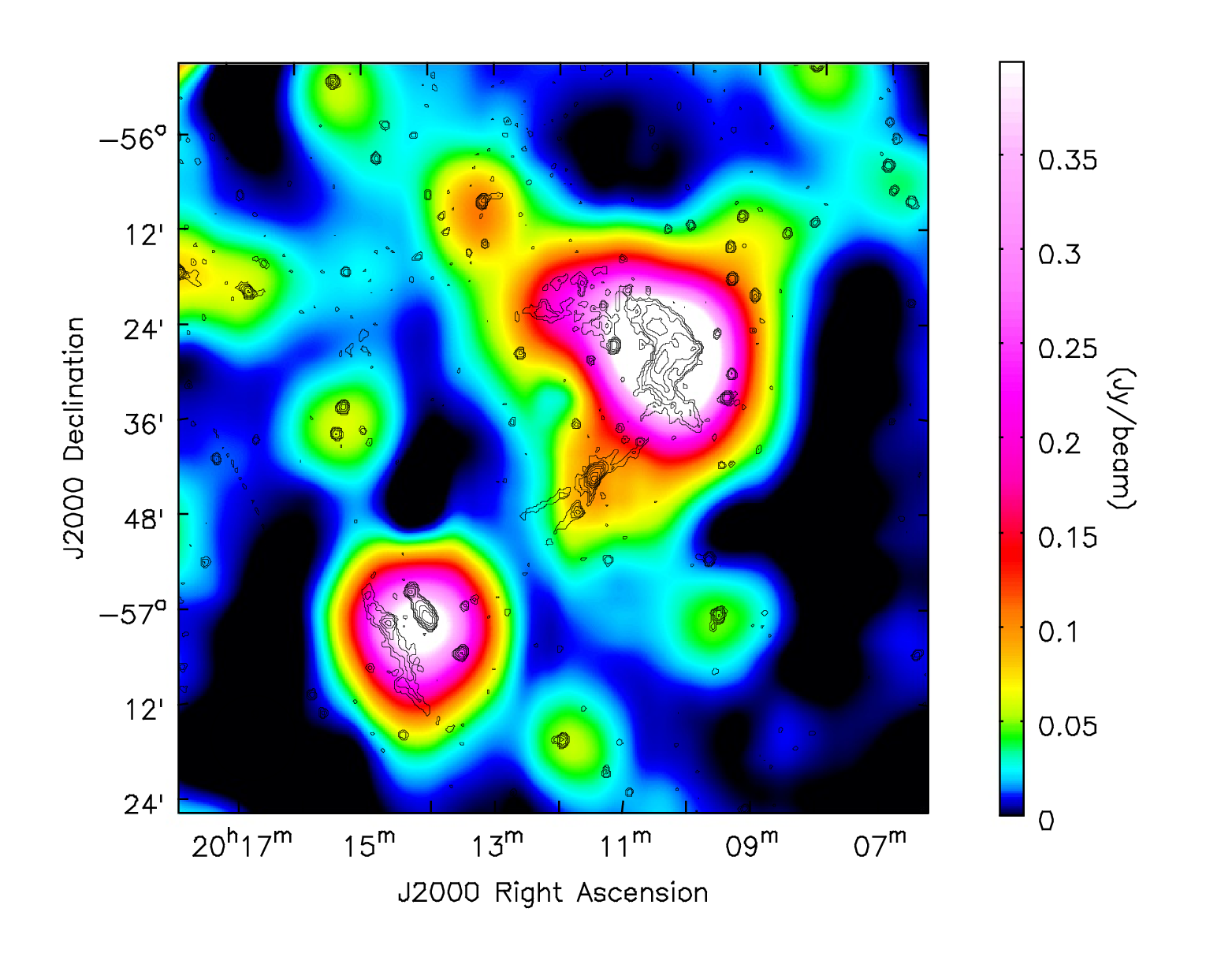}
\caption{{\bf Top--left:} Stokes I image of A3667 from S-PASS data taken at 2.3 GHz with
                the Parkes radio telescope. Compact sources have not been subtracted.
               Contour levels start from 30 mJy/beam and scale by a factor $\sqrt{2}$ thereafter.
               The noise budget is dominated by the confusion limit of 6 mJy. 
               Beam size is 10.75 arcmin and reported in the bottom-left corner. 
               {\bf Top--right:} Same image, but with the positions of the sources 
                detected with the ATCA observations marked (crosses). The area covered 
                by the ATCA  is also shown (black circle).
               {\bf Bottom--left:} Same as above, but with the ATCA sources subtracted, 
               including the head--tail radio galaxy
                B2007-569.  The centre of the cluster is marked with a black square 
                and major features
               described in the text are labelled.
               {\bf Bottom--right:} Same source--subtracted map as bottom--left panel,
              but with the contour levels of the 843 MHz
              SUMSS emission overlaid.  Levels start at 3~mJy/beam and scale 
               by a factor of 2 thereafter. Beam is 43"~$\times$~43"~cosec(Dec).  
                \label{a3667_spass:Fig}
}
\end{figure*}
 \begin{figure}
\centering
  \includegraphics[angle=0, width=1.0\hsize]{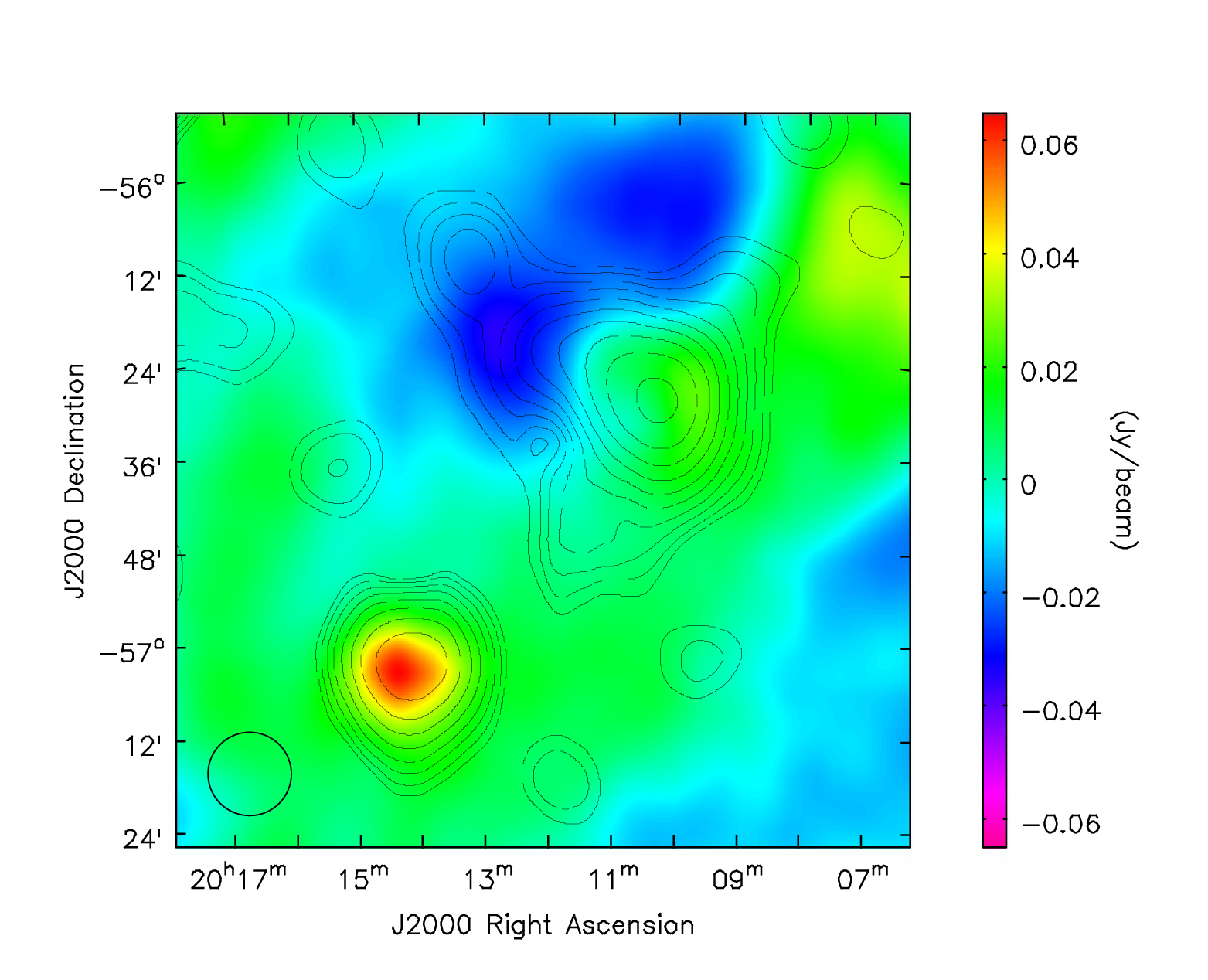}
  \includegraphics[angle=0, width=1.0\hsize]{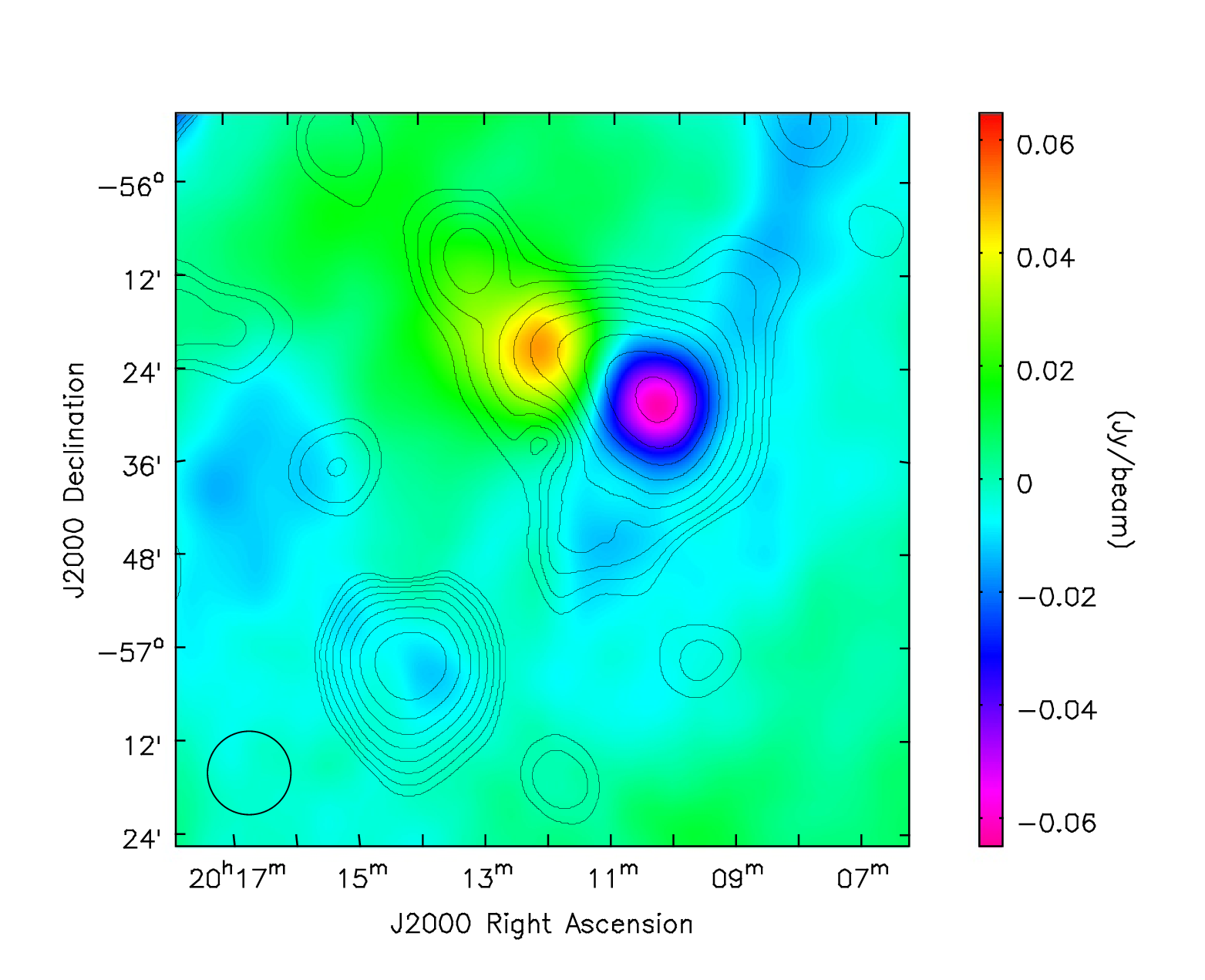}
\caption{Stokes Q (top) and U (bottom) image of A3667 from S-PASS data taken at 2.3 GHz with the Parkes radio telescope. Stokes I contour levels as from Figure~\ref{a3667_spass:Fig} are overlaid.  The beam size is shown in the bottom--left corner.
                \label{a3667_QU:Fig}
}
\end{figure}

Emission from the two relics is obvious at the NW and SE corner of the cluster and peaks at 360 and 300~mJy/beam in the 3.3~GHz Parkes data. 
However, no ICM emission from the central regions of the cluster is visible in the 843~MHz image. The head--tail galaxy B2007-569 can be seen to be associated with an X-shaped structure which is a SUMSS artifact not visible in other images of the same object at the same frequency (e.g. \citealt{rottgering97}).

Along with the two relics, our cleaned single-dish image shows a previously undetected radio bridge stretching for some 20' from the edge of the NW relic toward the central regions of the cluster. A second spot of diffuse emission can also clearly be seen associated with the peak of the cluster X-ray emission (see also Fig.~\ref{a3667_10cm_xray:Fig}). This halo seems to be connected to the second relic in the SE.  Because of the resolution it is not clear whether bridge, halo, and second relic are continuous or not. If so, this would be a huge structure stretching for 60-65', corresponding to 3.9-4.2 Mpc at the A3667 distance.

The cleaned emission from the bridge ranges from 28 to 35 mJy/beam (surface brightness of 0.14-0.18~$\mu$Jy/arcsec$^2$), a highly significant detection compared to the confusion limit. The integrated flux density is $102\pm8$~mJy.\footnote{The error includes the B2007-569 fitting error and confusion limit.}  The halo peaks at 22~mJy/beam with an integrated flux density of $44\pm6$~mJy. Though hints of halo emission have been seen in interferometric observations, these are likely due to imaging artifacts \citep{johnstonhollitt03}. This is the first clear detection of diffuse emission coincident with the central ICM of A3667.

Figure~\ref{a3667_spass:Fig}  shows the 2.3~GHz S-PASS total intensity map centred at A3667 before compact source subtraction along with the position of the sources detected with the ATCA observations and the ATCA field. The flux densities measured at 2.3~GHz are reported in Table~\ref{b2007_src:Tab}.

Figure~\ref{a3667_spass:Fig} also shows A3667 with  the ATCA field sources subtracted, including the head--tail galaxy (integrated flux density of 280~mJy).  The same image overlaid with the contour levels of the 843~MHz SUMSS data is also shown. 
Importantly, this cleaned single-dish image clearly shows the bridge stretching for 20-25' from the edge of the NW relic toward the central regions of the cluster.  The cleaned emission from the bridge ranges from $\sim$100 to $\sim$115~mJy/beam, a highly significant detection. This corresponds to a surface brightness of 0.21-0.25~$\mu$Jy/arcsec$^2$. The integrated emission is $160\pm8$~mJy.  The mean spectral index from 2.3 and 3.3 GHz is $\alpha_{\rm b}= -1.21\pm 0.15$. 
This further excludes the bridge being a residual contamination of the B2007-569 tail, whose spectrum is much steeper (spectral index between 1.4 and 2.3~GHz $\alpha_{\rm t}< -1.5$, \citealt{rottgering97}).

Emission from the NW and SE relics is also obvious and peaks at 
750 and 460~mJy/beam, respectively. 

The diffuse emission complex encompassing both the NW relic and 
the radio bridge covers a very extended area stretching for some 
40'$\times$40', that corresponds to $2.5\times2.5$~Mpc at the A3667 distance. 

Both relics show polarized emission in our 2.3 GHz data (Figure~\ref{a3667_QU:Fig}). 
The more extended NW relic shows polarized emission centred at the total intensity peak. 
 Its brightness is  64~mJy/beam for a polarization fraction of 8.5\%. 
 Previous observations at 1.4~GHz \citep{johnstonhollitt03} measured 10-40\% 
 fractional polarization across the relic, though this was at high resolution with 
 an interferometer.  Toward the NE it seems to blend with more extended and 
 weaker emission that does not appear to be associated with the cluster, and 
 is possibly due to Galactic foreground. This makes hard to measure the 
 total amount of polarized emission in the NW relic. 

The polarized component of the SE relic amounts to 63 mJy/beam, 
for a polarization fraction of 14\%. The polarization angle of both relics 
might be subject of some Faraday Rotation. Follow--up observations at other frequencies are required.

The radio bridge does not show obvious polarized emission. 
An excess of 4~mJy/beam (3.5\% polarization fraction) is measured 
compared to the surroundings, but this is likely mixed with the Galactic 
foreground and can be regarded as upper limit.

A small spur also seems to head NW from the NW relic, although four sources of 10-20~mJy at 843 MHz might account for this. The area was not covered in our ATCA follow up.

\section{Discussion}\label{disc:Sec}

\subsection{Radio interferometric data}
Even higher resolution radio data for the cluster are available from SUMSS (843 MHz, see Figures~\ref{a3667_10cm:Fig} and~\ref{a3667_spass:Fig}) and the ATCA~\citep{rottgering97}. The latter have been taken at 1.4 and 2.4~GHz, but only images at 1.4~GHz have been published. The two radio relics are obvious in those data sets and overlap well with our image.
There is no trace of the bridge, except in Molonglo Observatory Synthesis Telescope (MOST) images, where there are signs of a hint of emission too noisy to be significant \citep{rottgering97}.  The surface brightness of 0.25~$\mu$Jy/arcsec$^2$ we detect is weaker than that usually detectable with interferometers. In addition, the emission is probably smooth with most of its power at large angular scales where the interferometers have no sensitivity. 

Besides the bridge, the single--dish emission covers a more extended region than that seen by interferometer data, and results in a complex ``cloud" enclosing the NW relic and the bridge,
possibly stretching all the way down to the SE relic.

\subsection{Comparison with optical data}
\begin{figure}
\centering
  \includegraphics[angle=0, width=1.0\hsize]{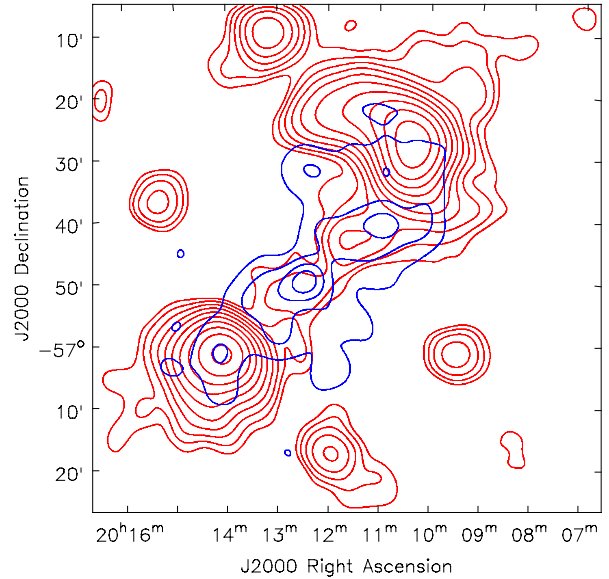}
\caption{Contour levels of galaxy number density (blue) along with those of 
                the 3.3~GHz source--cleaned emission of Figure~\ref{a3667_10cm:Fig} (red). 
               Levels are 40, 80, 120, and 160 galaxy/Mpc$^2$.  
               Galaxy positions and redshifts are from the 
               NASA Extragalactic Database (NED).
                \label{a3667_galden:Fig}
}
\end{figure}

Figure~\ref{a3667_galden:Fig} shows the galaxy number density of A3667. Galaxies have been selected from the NASA Extragalactic Database (NED) within a distance of one degree from the cluster centre and within a redshift  of $\Delta z = 0.01$. The latter allow separating cluster from background galaxies well \citep{owers09}. As found by \citet{owers09} there are two major peaks in the density distribution corresponding to the primary subcluster, near the A3667 centre, and to the secondary one just SE of the NW relic.  While the primary subcluster is almost at rest with the overall frame of A3667, the other one looks to be moving mostly on the plane of the sky at high speed (the average line-of-sight speed is already some 500~km/s). \citet{owers09} interpret this structure as a post major merger cluster, with the fast moving component heading NW following the NW relic shock front.

Compared to our images, the galaxy distribution looks well aligned with the radio bridge we have detected and with the feature connecting the two relics. In particular the moving cluster appears lie near the connection between bridge and relic. 
 There is clearly an association between the bridge and the direction of motion of both the outgoing relic  and the moving subcluster.

Emission is also clearly detected at the centre of the cluster at 3.3~GHz. 
Only a hint is visible at 2.3~GHz,
 but not enough to be significant possibly because of the higher confusion limit at this frequency.

\subsection{Comparison with X-ray data}

\begin{figure}
\centering
  \includegraphics[angle=0, width=1.0\hsize]{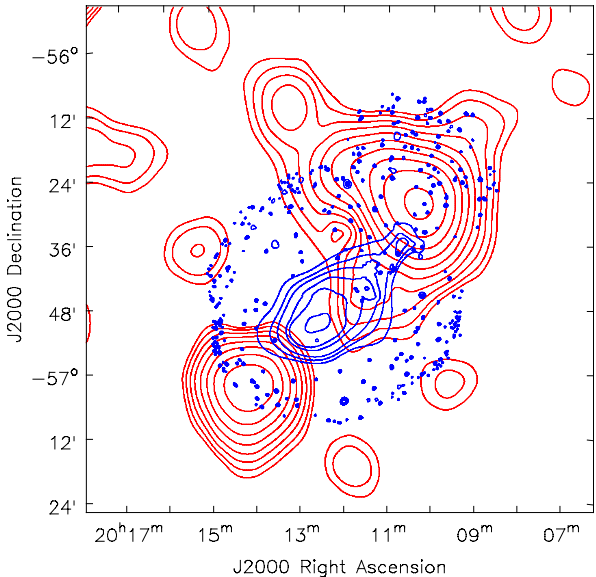}
\caption{Contour levels of X-ray emission (blue) from XMM data \citep{finoguenov10} along with those of  source-cleaned Stokes I emission of Figure~\ref{a3667_spass:Fig} (red).  X-ray levels are  1.6, 6.4, 12.5, 25, 50, 200x10$^{-6}$~cts/s/arcmin$^2$.
                \label{a3667_X:Fig}
}
\end{figure}

  Figures~\ref{a3667_10cm_xray:Fig} and~\ref{a3667_X:Fig} show the X-ray emission of A3667 detected by \citet{finoguenov10}  with XMM observations.  The brightest emission is from the centre of the cluster and an extended, smooth, and weaker emission  tail goes all the way to the NW relic slowly decreasing in brightness, stopping abruptly at its inner edge.

The radio bridge is well aligned with the X-ray tail. The two types of emission look clearly associated.  

A local X-ray maximum lies close to the NW end, offset compared to the position of the moving subcluster, even though \citet{finoguenov10} find them associated. However, our subcluster position is consistent with that of  \citet{owers09}. We do not find obvious association between the X-ray local peak and the moving subcluster.
  
\subsection{Interpretation} 
Several features distinguish the new {\it relic--bridge} from others seen: 1) this is one of the few double relic systems  (so shocks are likely propagating in the plane of the sky) with a bridge, removing uncertainties regarding projected radio halo structures that confuse other systems;  2) there is a coincident X-ray tail detected at high significance; 3) A3667 is the only bridge in which the relic is a confirmed shock \citep{finoguenov10}; 4) the dynamics of the merging clusters is known very well, so we know that the bridge is ``in the wake" of the passing shock.

The currently accepted  interpretation is that A3667 is a post major merger cluster that has generated two outgoing shock waves (the two relics).  Simulations and theoretical analysis predict that turbulence is generated in the post--shock region  \citep{kang07,paul11,vazza11}, but no clear evidence has yet been found. Turbulence scales with thermal energies \citep{vazza11}, so that it is expected to be traced by enhanced X-ray emission. In A3667, the X-ray emission tail is aligned with the assumed path of the outgoing shock. 
Further, we clearly detect a radio bridge connecting the cluster to the NW relic and which is also aligned with the path followed by the relic, the moving cluster, and the X-ray emission tail. We conclude that the synchrotron emission is correlated to the post--shock turbulence trailing the relic shock. 

  The exact coupling between the turbulence and synchrotron emission could take several forms. According to turbulence re--acceleration models (e.g. \citealt{brunetti01, petrosian01}), one would expect synchrotron emission to be generated in this area, assuming that the initial passage of the shock generated sufficient seed CRe$^{\pm}$. In this framework one would expect the radio emission to fade as the turbulence dissipates and the seed electrons age. The upper limit on the polarization  of the bridge is consistent with no polarization, which we would expect from synchrotron emission generated in a turbulent magnetic field. 
The time passed since the shock wave was at the current inner edge of the NW relic can be estimated from the relic thickness ($\sim$10') and 
the speed of the shock ($1210\pm220$~km/s, \citealt{finoguenov10}). The resulting 0.39--0.84~Gyr (2--$\sigma$ C.L.)
is consistent with the time required 
by the turbulent re-acceleration model to reaccelerate electrons ($\sim$0.4~Gyr with an upper limit of $\sim$0.7~Gyr, \citealt{cassano05}) 
and gives sufficient time to populate the space behind the relic with synchrotron emitting electrons.

The magnetic field intensity cannot be measured directly with the data available, but an estimate can be done using the equipartition magnetic field equation  from the emission intensity, spectral slope, and line-of-sight path length $l$ \citep{beck05}. From the 3.3~GHz image we measure a bridge width of 9' (beam-smearing corrected), which corresponds to 560~kpc. Assuming a cylindrical geometry (i.e. $l$ equal to the bridge width) we compute a equipartition magnetic field $B_{\rm eq} = 2.2\pm0.3$~$\mu$G. We have assumed a proton-to-electron number density ratio $K_0 = 200$.\footnote{Typical expected values in clusters are $K_0 = 100$--300 (e.g. see \citealt{pfrommer04}): The magnetic field estimate would vary from 1.8 to 2.4~$\mu$G within such a range.} 
  
An alternative explanation is that the bridge has a hadronic origin, resulting from CRp--p collisions and the resulting secondary CRe$^{\pm}$. The presence of a bridge but not a (strong) radio halo is troubling for this model, but could be due to freshly accelerated CRp (at the shock) interacting with a magnetic field strongly amplified by the post--shock turbulence (e.g. Keshet 2010), where the field may have already dissipated in the central regions. Detailed spectral-index measurements may constrain the two models.

Other interpretations for the seeding of the CRe and the generation of the turbulence are also possible, although less convincing. The relativistic electrons might in fact have been seeded by the head--tail galaxy B2007-569. However,  the direction of the tails, 
mostly from N to N--NW 
(see also Figure~2 in \citealt{rottgering97}), 
cannot account for the direction of the bridge, whose axis is NW to W--NW.
A second alternate interpretation is that the radio emission is still generated by turbulence, but caused by the motion of the fast moving subcluster. However, its position in the middle of the X-ray tail cannot account for the X-ray emission of the section between subcluster and NW relic.

In all cases, the indicators point toward the large-scale synchrotron emission being associated with post--shock turbulence generated by a major merger in a massive cluster. Although predicted by simulations  \citep{kang07,paul11,vazza11}, this is the first time such emission is detected with high significance and clearly associated with the path of a confirmed shock. This result supports a turbulent re-acceleration model for the relativistic electrons, that naturally explains the presence of a synchrotron bridge in the post--shock region.
 
 The radio luminosity of the halo is $P_{1.4} = 7.5 \times 10^{23}$~W/Hz, extrapolated to
1.4~GHz using a spectral index of -1. With A3667's X-ray
luminosity ($L_X = 9.3\times10^{44}$~erg/s, \citealt{reiprich02}), this sets the halo between the
$P_{1.4}$--$L_X$ correlation and the {\it off-state} found in \citet{brown11b}.

 \section{Summary} We have detected a radio bridge of unpolarized synchrotron emission connecting the NW relic of Abell 3667 to the central regions of the cluster. This emission is further aligned with a diffuse X-ray tail, and represents the most compelling evidence for an association between ICM turbulence and diffuse synchrotron emission. Though the origin of the relativistic electrons is still unknown, the turbulent re-acceleration model \citep{brunetti01, petrosian01} provides a natural explanation for the large-scale emission. Further spectral and high-resolution observations are needed in order to differentiate this from potential hadronic secondary models of cosmic-ray acceleration in the post--shock region. We further detect diffuse emission coincident with the central regions of the cluster for the first time.
 
\section*{Acknowledgments}

This work has been carried out in the framework of the S-band All Sky Survey collaboration (S-PASS).
We would like to thank Alexis Finoguenov for the electronic image of the XMM X-ray emission of A3667 and an anonymous referee for useful comments that helped improve the paper.  
B.M.G. acknowledges the support of an Australian Laureate Fellowship from the Australian Research Council through grant FL100100114.
M.H. acknowledges the support of the research programme 639.042.915, which is partly financed by the Netherlands Organisation for Scientific Research (NWO).
The Parkes radio telescope is part of the Australia Telescope National Facility which is funded by the Commonwealth of Australia for operation as a National Facility managed by CSIRO. 
The Australia Telescope Compact Array is part of the Australia Telescope National Facility which is funded by the Commonwealth of Australia for operation as a National Facility managed by CSIRO. We acknowledge the use of NASA Extragalactic Database.

\bsp

\label{lastpage}

\end{document}